\documentclass[runningheads]{llncs}
\usepackage[T1]{fontenc}
\usepackage{graphicx}
\usepackage{hyperref}
\usepackage{color}
\usepackage{amsmath}
\usepackage{booktabs}
\usepackage{multirow}

\usepackage{tablefootnote}
\usepackage{adjustbox} 
\usepackage{tabularx}
\newcolumntype{Y}{>{\centering\arraybackslash}X}
\usepackage{bbold} 
\usepackage{arydshln} 
\usepackage{caption} 
\usepackage{cleveref} 
\usepackage{makecell} 

\begin{document}

\title{Mitigating False Predictions In Unreasonable Body Regions}
%
%


\author{Constantin Ulrich*\inst{1,4,5} \and
Catherine Knobloch* \inst{1} \and
Julius C. Holzschuh \inst{1} \and
Tassilo Wald\inst{1,2,6} \and
Maximilian R. Rokuss \inst{1,6} \and
Maximilian Zenk\inst{1,5} \and
Maximilian Fischer\inst{1} \and
Michael Baumgartner\inst{1,2,6} \and
Fabian Isensee\inst{1,2} \and
Klaus H. Maier-Hein\inst{1,3}}


\authorrunning{C. Ulrich et al.}

\institute{German Cancer Research Center (DKFZ), Heidelberg, Division of Medical Image Computing, Germany\\
\and Helmholtz Imaging, DKFZ, Heidelberg, Germany\\
\and Pattern Analysis and Learning Group, Department of Radiation Oncology, Heidelberg University Hospital, Heidelberg, Germany \\
\and National Center for Tumor Diseases (NCT), NCT Heidelberg, A partnership between DKFZ and University Medical Center Heidelberg\\
\and Medical Faculty Heidelberg, University of Heidelberg, Heidelberg, Germany
\and Faculty of Mathematics and Computer Science, Heidelberg University, Germany
\email{constantin.ulrich@dkfz-heidelberg.de}}
\maketitle 
\begin{abstract}
Despite considerable strides in developing deep learning models for 3D medical image segmentation, the challenge of effectively generalizing across diverse image distributions persists. While domain generalization is acknowledged as vital for robust application in clinical settings, the challenges stemming from training with a limited Field of View (FOV) remain unaddressed. This limitation leads to false predictions when applied to body regions beyond the FOV of the training data. 
In response to this problem, we propose a novel loss function that penalizes predictions in implausible body regions, applicable in both single-dataset and multi-dataset training schemes. It is realized with a Body Part Regression model that generates axial slice positional scores. Through comprehensive evaluation using a test set featuring varying FOVs, our approach demonstrates remarkable improvements in generalization capabilities. It effectively mitigates false positive tumor predictions up to 85\% and significantly enhances overall segmentation performance.

\end{abstract}
\def\thefootnote{*}\footnotetext{Equal Contribution.}\def\thefootnote{\arabic{footnote}}

\section{Introduction}

\begin{figure}[h]
\centering
\includegraphics[width=1\textwidth]{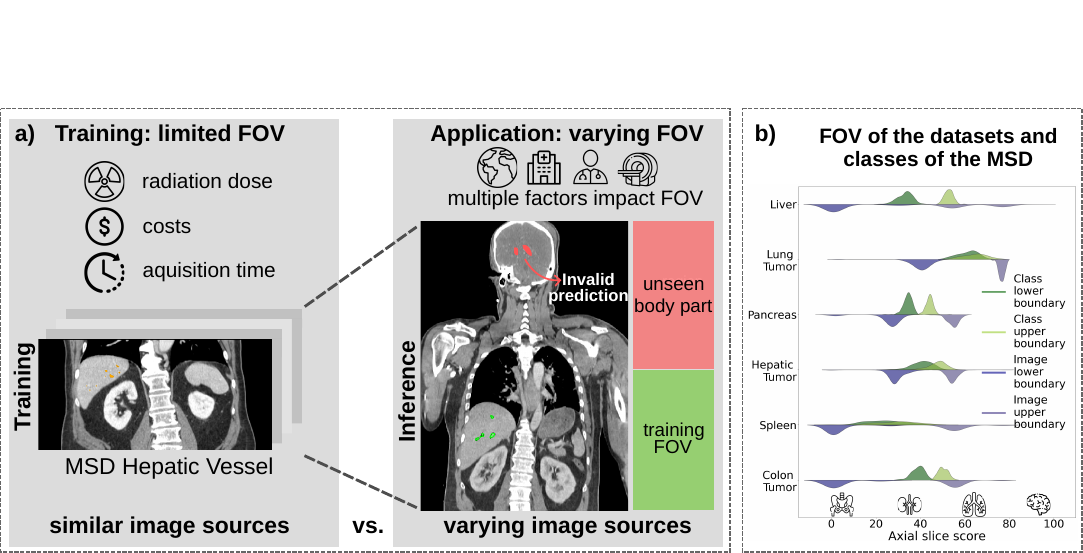}
\caption{a) shows an example of a training dataset with a limited Fiel of View (FOV) caused by the desire to reduce radiation dose, scanning time and costs. Typically, the final model is then applied to images from different sources and studies with varying FOVs and tends to make anatomical implausible predictions in unseen body parts. b) shows the FOV of the datasets of the Medical Decathlon as well as the distribution of the location of the upper and lower position of the target structure, which was determined using a Body Part Regression model.} \label{overview_figure}
\end{figure}

In modern healthcare, 3D medical image segmentation is crucial to support clinicians in treatment planning, disease monitoring, and anatomical as well as pathological structure analysis. Despite significant advancements of deep learning models for medical image segmentation \cite{isensee_nnu-net_2021,ronneberger_u-net_2015,huang2023stunet,roy,extending}, a persistent challenge remains: a robust generalization to image distribution shifts. Many existing methodologies have attempted to address this by focusing on improving generalization between clinics, scanner types, imaging protocols, and demographics \cite{seg_generalization,He_Yang_Li_Li_Chang_Yu_2019,Kloenne2020,GeneralizingZhang}. However, a fundamental aspect often overlooked is the limited Field of View (FOV) inherent in most 3D medical image datasets used for model training. This restricted FOV primarily arises due to constraints imposed by factors such as the reduction of radiation dose, acquisition time, and associated costs, especially in datasets restricted to certain pathologies of interest. Even if the images of a public dataset come from different sources with originally larger and more diverse FOVs, the images are often cropped during the manual annotation and publishing process to streamline image processing and minimize memory consumption. However, during inference, this results in tangible scenarios in which trained models are applied to images with a different FOV. This is a particularly severe problem when large image cohorts need to be automatically delineated. \\
The disparity between the restricted FOV during training and larger and more diverse FOVs encountered in real-world applications poses a major challenge to the adaptability and generalizability of segmentation models. As illustrated in \cref{overview_figure} a), models tend to make incorrect predictions in unseen regions not represented in the model's training data e.g. entail liver vessels in the head. Such false predictions are completely irrational for humans and these conspicuous errors not only diminish the measurable performance of these models but also drastically undermine the trust that patients and clinicians place in these methods. \\ 
In contrast to previous methods that addressed false predictions by an error prone additional post-processing step \cite{ji2023continual,heller2020kits19,bpregsarah}, we introduce a novel training approach designed to mitigate wrong predictions in body parts without label information. A novel Region Loss directly penalizes predictions in previously defined unreasonable body regions during model training. To determine the corresponding body regions, we employ a self-supervised trained Body Part Regression (BPR) model, which assigns each axial CT image slice a specific value corresponding to its axial position, ensuring that anatomically equivalent body parts are assigned similar values across patients and images \cite{bpregsarah,yan2018unsupervised}.\\
We evaluate the proposed method in both, a standard single-dataset (SD) training scenario as well as a multi-dataset (MD) training scenario, where multiple datasets with various FOVs and heterogenous class annotations are used concurrently \cite{multitalent}. 
Importantly, as we illustrate the potential of our method for 3D image segmentation in CT images, its versatility and the self-supervised training of the BPR model pave the way for straightforward expansion to a range of other 3D image modalities and tasks, including detection.
Our results demonstrate that the proposed method effectively allows generalization between varying FOVs and mitigates false positive predictions in anatomically unreasonable body parts. The code and the models will be published at: tba.

\section{Method}
In the following, we present our approach to prevent predictions in unreasonable body parts using a novel Region Loss. The method is implemented both in a MD and in a SD setting in conjuction with an unlabeled support dataset to provide the network with varying FOVs.

\subsection{Body part Regression Model}
To translate each anatomical body part of a CT slice into a standardized machine-interpretable position we employed the publicly available BPR model by Schuhegger et al. which is an extension of the work of Yan et al. \cite{bpregsarah,yan2018unsupervised}. While the training details can be found in the corresponding publications, the final model $f_\theta$ with parameters $\theta$ has learned a mapping of an axial slice $\hat x$ of a CT image $X$ to a slice score $s$ where the start of the pelvis is mapped to 0 and the head to 100: $f_\theta : \mathbb{R}^{2}\xrightarrow{} \mathbb{R}: f_\theta(\hat x) = s$.
The BPR model was applied to all training images to assign each image voxel $x_n$ in each slice to the corresponding encoded axial position $s_n$. 
We approximated the BPR scores for slices below the pelvis using linear extrapolation.  Fig. \ref{overview_figure} (b) shows the image FOV, as well as the foreground class distributions of all CT datasets of the Medical Segmentation Decathlon \cite{Antonelli2022}). None of the datasets covers all body parts from the pelvis to the head and some datasets exclusively cover a FOV limited to the extent of the target structure.

\subsection{Region Loss}
For training the segmentation networks, we introduce a novel loss function to penalize predictions in unreasonable body parts. The loss is based on the weighted Binary Cross Entropy (wBCE) loss defined for the network prediction $\hat y$ and the corresponding binary ground truth $y$ of an individual voxel as follows:
\begin{equation}
    wBCE(\hat y_n, y_n, w_n) = w_n[y_n\log(\hat y_n) + (1-y_n)\log(1-\hat y_n)] = w_n \log(1-\hat y_n)  = l_n, 
\end{equation}
where $w_n$ is a weighting factor for each predicted voxel $\hat y_n$ with $n \in [1,N]$ where N is the number of voxels in an image. This factor is used to determine in which body part the loss term penalizes the predictions and in which body part predictions were considered to be valid. Within the invalid region, the ground truth of a target class is expected to be zero, which translates to $y_n = 0$. In \cref{boundaries}, it is explained how the upper and lower boundaries for the valid body regions $S^{max}, S^{min}$ were determined. Taking into account the uncertainties of the BPR mapping and variations in the human anatomy, we decided to choose a smooth transition of the voxel weighting between valid and invalid regions using a Gaussian smoothing:
\begin{equation}\label{eq:weights}
    w_n(s_n, S^{min}, S^{max}) = \begin{cases}
        0 & :S^{min}\leq s_{n}\leq S^{max} \\
        1 - \exp(\frac{-(s_{n}-S^{min})^2}{2(\sigma^{min})^2}) & :s_{n}<S^{min} \\
        1 - \exp(\frac{-(s_{n}-S^{max})^2}{2(\sigma^{max})^2}) & :s_{n}>S^{max},
    \end{cases}
\end{equation}
where $\sigma^{max}$ and $\sigma^{min}$ correspond to the standard deviation of the distribution of all minimum and maximum slice scores of a target class over all training images. \\
The final Region Loss only penalizes the largest k percent of the wBCE loss values of all predictions \cite{Lyu_2022} made in the invalid regions. Otherwise, many small predicted probabilities would reduce the overall loss signal:
\begin{equation}\label{eq:topk_formulated}
   RL = TopK(L) = \alpha \frac{1}{M}\sum_{n=1}^Ml_n,
\end{equation}
where $\alpha$ is a weighting factor, $M=\hat N\cdot\frac{k}{100}$, $L=\{l_n|\forall n \in [1,\hat N]\}$, $\hat N$ is the number of voxels within the defined invalid region, and $l_n$ is the $n$-largest wBCE voxel value.

\subsection{Experiments}
\subsubsection{Multi-dataset training}
In this study, we employ the loss function described above within the recently introduced MultiTalent framework for MD training \cite{multitalent}. In this MD setting, a network undergoes training using multiple datasets, each containing a distinct subset of annotated target structures. With the substantial expenses associated with annotating 3D medical images, the scientific community has recently intensified its focus on the challenge of learning from partially labeled datasets \cite{FLARE22}. MultiTalent comes with a dataset and target adaptive loss function to calculate the loss solely for the classes that were annotated in the corresponding dataset, while simultaneously predicting all classes from all datasets \cite{multitalent}. Consequently, the network encounters FOVs where ground truth annotations for certain classes are absent. For these classes, we incorporate the Region Loss penalize predictions in unreasonable body regions. The total loss can be formulated as follows:

\begin{equation}
    L^{(k)} = \sum_{c^{(k)}}L_c + \sum_{c^{(\neg k)}}RL_c
\end{equation}
where we add up the class and adaptive loss $L$ over all annotated classes $c^{(k)} \in C^{(k)}$ of a dataset $k \in K$ and the Region Loss $RL$ over all unlabeled classes $c^{(\neg k)} \in C^{(\neg k)}$, which corresponds to all classes that are annotated in the remaining datasets of $K\setminus k$. 
$K$ denoted the collection of datasets used for training the MultiTalent network. The MultiTalent network was trained on 13 datasets as the original work \cite{antonelli2021medical,BTCV,BTCV2,structseg,lambert2019segthor,roth2015deeporgan,clark_cancer_2013,NHpancreas,heller2020kits19}. We restricted the evaluation to the classes of 5 Medical Segmentation Decathlon (MSD) CT datasets \cite{Antonelli2022}.

\subsubsection{Single-dataset training}
In a conventional training scenario, where a network is trained using a SD containing a specific set of annotated target structures, the network does not encounter FOVs lacking ground truth annotations. Consequently, we chose 120 CT images from the AutoPET2 dataset \cite{autopet} that do not contain annotations of the target classes but possesses a distinct and larger FOV. During training, we select these images with a probability of $p$. For all annotated images from the SD, the default nnU-Net supervised loss was computed \cite{isensee_nnu-net_2021}, while for images lacking label information, the Region Loss was employed. In this work, we tested the proposed method for 5 MSD CT datasets \cite{Antonelli2022}.

\subsubsection{Implementation}
All experiments conducted in this study utilize the established nnU-Net framework \cite{isensee_nnu-net_2021}. The slice scores are obtained by employing the BPR model developed by Schuhegger et al. prior to training the segmentation model to mitigate computational overhead.
A three-dimensional array, which has the same shape as the raw data and contains the slice scores for each corresponding voxel at the same position, is generated after applying the BPR model to the image. All spatial preprocessing steps and augmentations such as rotation, translation, and mirroring, are applied to both the training data as well as the slice score arrays. During training, these arrays are utilized to determine the anatomical region of the input patches in which the Region Loss penalizes predictions.
While many design decisions align with the default suggestions provided by nnU-Net and MultiTalent, an overview of all altered hyperparameters can be found in the appendix in \cref{hyperparameters}.

\subsubsection{Defining valid regions}
\label{boundaries}
 To find the best boundaries that delineate the anatomical regions where predictions should be penalized by Region Loss, multiple MultiTalent networks were trained with incrementally increasing valid regions. Initially, we chose the mean value of the minimum and maximum slice scores from the distribution corresponding to the lowest and highest ground truth label positions for each class. Because the distributions' positions and widths differ across classes, we extended the valid region step-wise according to the standard deviation of the minimum and maximum slice score distribution. The outcomes for all classes within the MultiTalent dataset collection are shown in the appendix in \cref{all_boundaries_Dice_scores}. Given the variations of the best boundary definition across classes based on the cross-validation results, we selected the best boundaries for each class for all subsequent experiments.

\subsubsection{Baselines}
For both the MD and SD training settings, we employ the default training approach alongside two additional postprocessing schemes as baselines. The first postprocessing baseline utilizes the BPR model to retrospectively crop all predictions outside of the valid regions (BPR postprocessing). Consequently, the BPR model must first be applied to all test images to create a mask for the valid regions. The boundaries for this postprocessing cropping were determined through optimization on the validation predictions of the MD training and the SD training. In the second postprocessing baseline, all connected components except for the largest component LC were removed (LC postprocessing). This approach is particularly applicable in organ segmentation tasks where only one target instance is typically expected, making it a well-established baseline. Our method stands as the first approach that addresses the FOV issue directly during the training not relying on an additional postprocessing step.

\section{Results}
\label{res_section}
\begin{figure}
\centering
\includegraphics[width=1\textwidth]{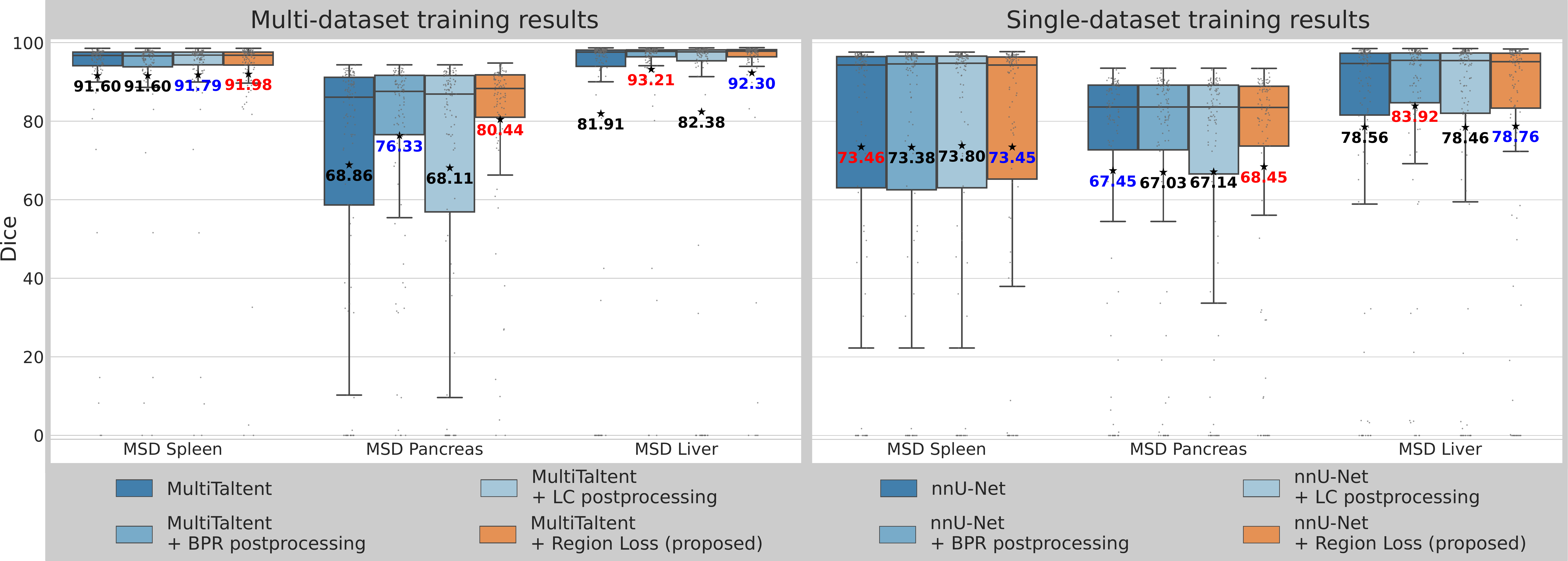}
\caption{The proposed Region Loss achieves always the best (red) or second best (blue) mean Dice. Additionally, the Region Loss improves the boundaries of the lower two boxplot quartiles compared to default MultiTalent or default nnU-Net. This indicates, that the Region Loss succesfully improves inferior cases.}\label{allres}
\end{figure}
To evaluate the efficacy of the proposed method, we used all 146 images from the official test and validation set of the \textit{TotalsegmentatorV2} dataset \cite{totalseg}. This dataset includes comprehensive field-of-views (FOVs) and corresponding organ annotations, akin to the annotations found in the MSD datasets.
In \cref{allres} on the left, we present the results of the MD training setting. Notably, the proposed Region Loss consistently resulted in a better performance compared to the default MultiTalent baseline across all datasets. The mean Dice score has not only improved, but the narrower shape of the performance distributions, visualized by the boxplots, also suggests an enhanced consistency in the results with fewer negative outliers. 
Adding a BPR postprocessing step enhanced pancreas and in particular liver segmentation. The boundaries for cropping the predictions in this additional postprocessing step were optimized on the 5-fold cross-validation results. 
When the unknown inference data distribution resembles the training data, this postprocessing baseline serves as an upper limit for our method. Conversely, applying BPR postprocessing adversely impacted spleen segmentation, resulting in the removal of some true positive predictions. In general, this postprocessing heavily relies on the accuracy of the BPR model and the selection of optimal boundaries, which is challenging without access to the inference data. Removing all but the largest component only resulted in a notable positive impact on spleen segmentation. \\
In the SD results presented in \cref{allres} on the right, the overall trend remains consistent, and the proposed method shows a slight improvement for all organs compared to default nnU-Net training.
It is noteworthy that the MD results are significantly better than the SD results. While the initial MultiTalent publication hinted at a modest advantage over SD training within in-distribution settings, this out-of-distribution scenario underscores MultiTalent's superior generalization capabilities. Upon manual examination of the SD predictions, three primary cases in which failure predictions occur became apparent: highly noisy CT images, FOVs containing only small portions of the target structures and pathologies in the target structure (see appendix \cref{Example}. While the MD collection may include similar cases, the SDs cover a less diverse image distribution.
As a result, the SD models have larger variability and perform exceptionally poorly in some rare cases, which may overshadow the impact of the proposed method. For a further perspective of the outcomes, a stability ranking analysis is presented in the appendix in \cref{ranking} \cite{Wiesenfarth2021}. \\
Unfortunately, no corresponding ground truth annotations for the tumor classes present in the MSD datasets were available in the test dataset, despite the presence of images with pathologies. Therefore, a clinician carefully inspected all tumor predictions with a total volume of more than approximately $1cm^3$ to evaluate the effectiveness of our method. The results are presented in  \cref{fig:FPbars} for the MD setting and in \cref{tab_datasets_res} in the appendix. In \cref{fig:FPbars}, it can be seen that the proposed region loss successfully mitigates tumor predictions in anatomical spurious body regions for the MD training setting to 85.7\%. Table \cref{tab_datasets_res} in the appendix shows that MultiTalent finds more True Positives (TP) but also more False Positives (FP) tumors than nnU-Net. Again, we believe that MultiTalent performs better due to the MD training, but the \textit{Sigmoid} activation function allows more FP predictions than the \textit{Softmax} function used by default for nnU-Net. Most remaining FP predictions correspond to anatomical regions within the valid axial region, such as false liver tumor predictions in the spleen. In the SD setting, the region loss has no observable influence on reducing the FP predictions, but training with the unlabeled support set dataset seems to have a positive influence on finding more true positive tumors in the relevant regions. We will discuss this more thoroughly in the next section. As it was ambiguous to determine TP and TR for the Colon MSD task, due to the difficulty of the task itself, we have omitted this dataset \cite{Antonelli2022}.
\begin{figure}
    \centering
    \includegraphics[width=0.8\textwidth]{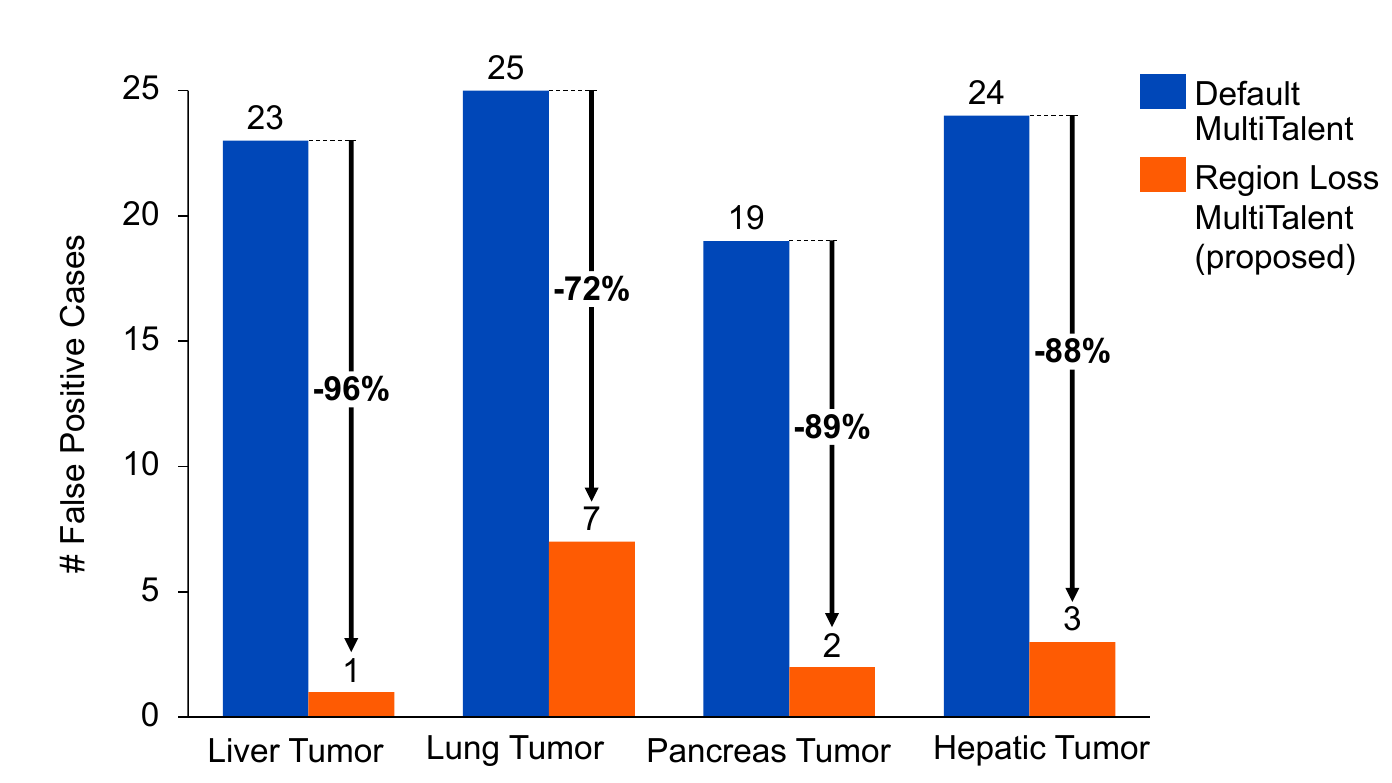}
    \caption{The proposed Region Loss mitigates most anatomical implausible False Positive tumor predictions in the MD setting. }
    \label{fig:FPbars}
\end{figure}

\section{Discussion}
The paper presents a novel method aimed at enhancing generalization across varying FOVs where supervision was initially not feasible using the available training data. This innovative approach utilizes a BPR model to encode axial slice scores, enabling the identification of anatomical regions where the introduced Region Loss function penalizes implausible predictions. Our method demonstrates superior performance in both MD and SD training settings, effectively reducing false positive predictions in anatomically unreasonable regions. \\
In contrast to post-processing methods, our approach offers greater stability, as post-processing techniques may inadvertently remove true positive predictions and rely on an additional error-prone step. Indeed, we encountered situations where manual correction of the slice scores was required while applying the BPR model to the images. 
In general, our results underscore the robustness of MultiTalent, particularly attributable to its MD training approach, compared to SD models. While MultiTalent's use of the \textit{Sigmoid} activation function may lead to more false positive predictions, our proposed method effectively mitigates this issue. 
The significance of the results in the SD setting was limited. Including an additional validation set would enable us to fine-tune our method's hyperparameters more effectively. However, it is challenging to find datasets with matching annotated target structures and an extended FOV. Additionally, because the unlabeled support set came from a single data source, the models may just learn to ignore predictions for images within that distribution. As part of our future work, we aim to integrate a more diverse support set, akin to the MD framework, which is expected to improve the efficiency of our method.
We have evaluated the efficacy of our method for segmentation in CT images, but it can potentially be extended to all 3D modalities and tasks such as detection.

\clearpage

\bibliographystyle{splncs04}
\bibliography{bibfile.bib}

\clearpage

\section*{Appendix}

\begin{figure}[h!]
    \centering
    \includegraphics[width=0.9\textwidth]{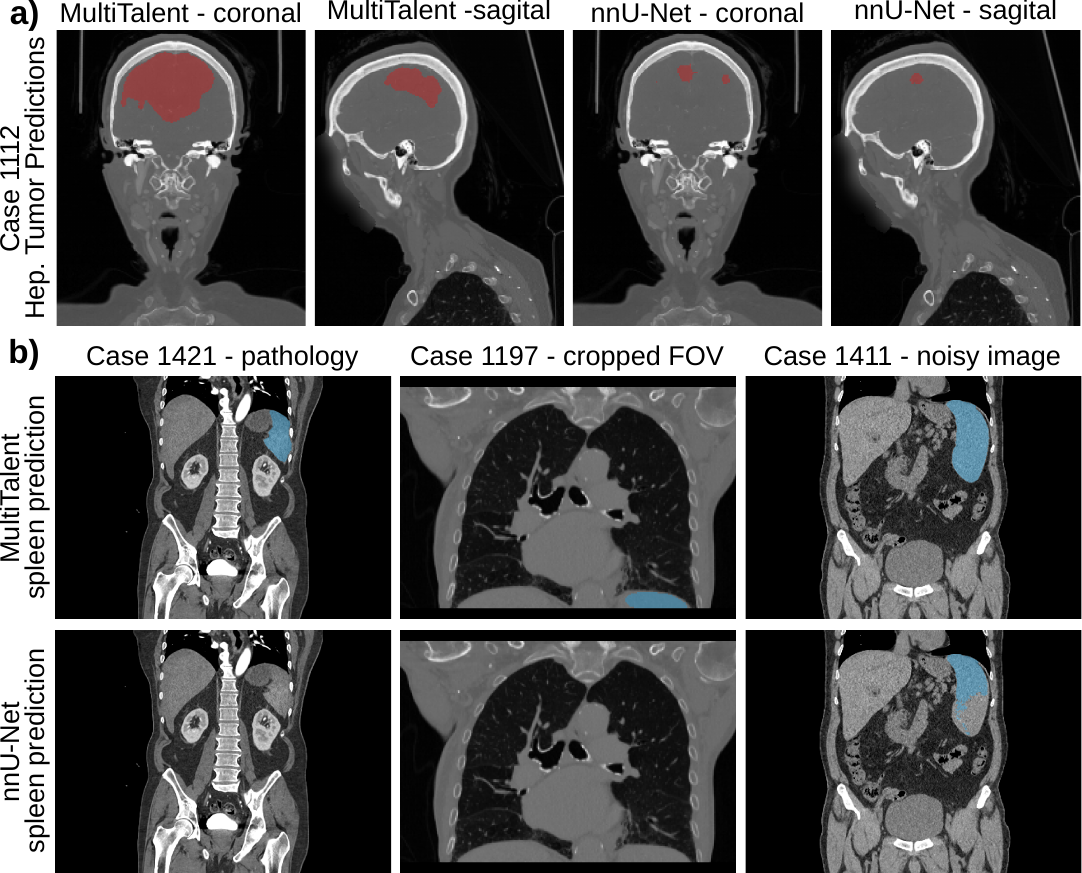}
    \caption{a) An example of unreasonable hepatic liver tumor predictions of MultiTalent and nnU-Net in the head that our proposed Region Loss completely mitigates. b) Example spleen predictions for which MultiTalent is more robust compared to nnU-Net.} \label{Example}
    \includegraphics[width=0.9\textwidth]{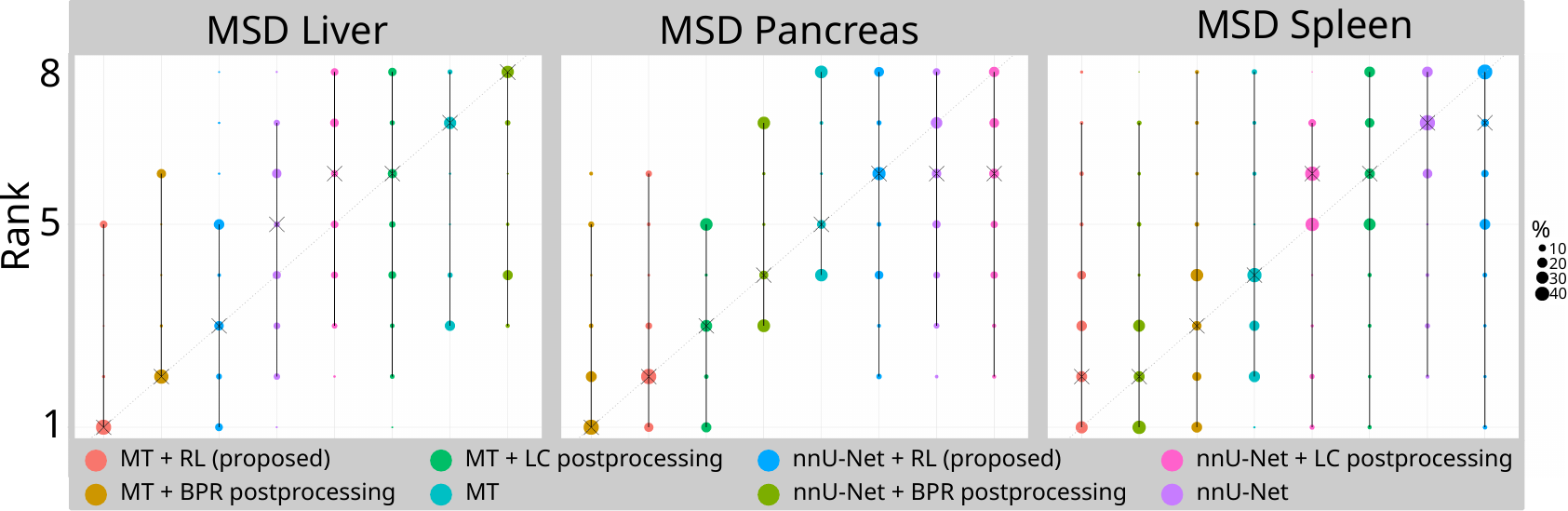}
        \caption{Comparison of the Ranking stability of all methods following Wiesenfarth et al. \cite{Wiesenfarth2021}. The area of each blob is proportional to the relative frequency of the achieved rank across 1000 bootstrap samples. The median rank for each algorithm is indicated by a black cross. 95\% bootstrap intervals across bootstrap samples are indicated by black lines.} \label{ranking}
\end{figure}

\begin{table}[t]
\centering
        \caption{List of all MSD 5-fold cross-validation Dice scores for the MultiTalent networks trained with different valid boundaries between $\mu$ and $\mu\pm6\sigma$ to determine the best boundaries for each class. The networks were trained for 1000 epochs to reduce compute overhead and without a Gaussian boundary weighting to find the best-fitting boundaries for each class. }\label{all_boundaries_Dice_scores}
    \begin{adjustbox}{width = 0.9\textwidth}
    \begin{tabular}{cccccccccccccc}
    \toprule
        \multicolumn{2}{c}{class} & \multicolumn{2}{c}{mean slice scores} &\multicolumn{10}{c}{Dice for different offsets to the mean slice scores }\\
        \cmidrule(lr){1-2}\cmidrule(lr){3-4}\cmidrule(lr){5-14}
         no & name & minimum & maximum& $0\sigma$ & $0.5\sigma$ & $1\sigma$ & $1.5\sigma$ & $2\sigma$ & $2.5\sigma$ & $3\sigma$ & $4\sigma$ & $5\sigma$ & $6\sigma$\\
        \midrule
        1 & liver & $33.09\pm3.31$ & $52.73\pm1.88$  & 95.72 & 95.80 & 95.82 & 95.79 & \textcolor{blue}{95.89} & 95.84 & 95.85 & 95.84 & 95.89 & 95.81 \\ 
        2 & liver tumor & $39.4\pm5.46$ & $48.49\pm4.39$  & \textcolor{blue}{64.25} & 62.55 & 63.55 & 62.79 & 63.15 & 62.82 & 64.23 & 61.99 & 62.13 & 62.59 \\ 
        3 & lung tumor & $61.25\pm4.5$ & $64.55\pm5.15$  & 67.23 & 67.88 & 65.95 & 67.54 & 67.35 & 65.43 & 66.14 & 68.02 & \textcolor{blue}{68.60} & 66.68 \\ 
         4 & pancreas & $34.34\pm1.96$ & $44.1\pm1.68$  & 79.86 & 79.97 & 79.97 & 80.01 & \textcolor{blue}{80.03} & 79.85 & 79.90 & 79.78 & 79.81 & 80.02 \\ 
        5 & pancreas cancer & $37.67\pm2.74$ & $40.38\pm2.88$  & 55.29 & 54.51 & 55.07 & \textcolor{blue}{55.84} & 55.62 & 55.32 & 55.76 & 55.11 & 55.27 & 55.56 \\ 
        6 & hepatic vessel & $36.14\pm4.09$ & $51.61\pm1.74$  & 63.76 & \textcolor{blue}{64.17} & 64.14 & 63.92 & 63.76 & 64.07 & 63.89 & 64.00 & 64.01 & 63.89 \\ 
        7 & liver cancer & $40.74\pm4.66$ & $47.65\pm3.96$ & 70.90 & \textcolor{blue}{70.95} & 70.27 & 70.71 & 70.92 & 70.78 & 70.55 & 69.75 & 69.93 & 70.66 \\ 
        8 & spleen & $39.87\pm2.34$ & $50.37\pm1.93$  & 96.13 & 96.07 & 96.14 & 96.26 & 96.29 & 96.26 & 96.20 & 96.23 & \textcolor{blue}{96.30} & 96.27 \\ 
        9 & colon cancer & $24.03\pm8.83$ & $29.49\pm9.29$  & 52.96 & 52.41 & 51.91 & 51.21 & \textcolor{blue}{53.10} & 51.31 & 50.96 & 52.42 & 51.04 & 49.79 \\ 
        \bottomrule
    \end{tabular}
    \end{adjustbox}

\centering
\caption{Overview of the training hyperparameters. For the MD training, the target spacing, patch size, and number of epochs were fixed as in the original publication \cite{multitalent}. Additionally, each dataset is sampled with a probability that is inversely proportional to $\sqrt{n}$, where $n$ is the number of images per dataset. For all experiments, we selected a batch size of two. For the SD training, we randomly selected unlabeled images with a 20\% probability and increased the default nnU-Net number of training epochs accordingly.} \label{hyperparameters}
\begin{adjustbox}{width=0.8\textwidth}
\begin{tabular}{lcccccccc}
\toprule
Framework & Setting & Batch size & Spacing & Patch size & Sampling prob. & epochs &  $\alpha$ & K \\
\midrule
MultiTalent & MD & 2& [1.5,1,1] & [96,192, 192] & default & 2000  & 1 & 1\\ 
nnU-Net  & MSD Liver & 2& [1 ,0.77 , 0.77 ] & [128, 128, 128] & 0.2 & 1200  & 10 & 1\\ 
nnU-Net  & MSD Lung & 2& [1.24, 0.79, 0.79] & [80, 192, 160] & 0.2 & 1200  & 10 & 1\\ 
nnU-Net  & MSD Pancreas & 2& [2.5, 0.80, 0.80] & [40, 224, 224] & 0.2 & 1200  & 10 & 1\\ 
nnU-Net  & MSD H.Vessel & 2& [1.50, 0.80, 0.80] & [64, 192, 192] & 0.2 & 1200  & 10 & 1\\ 
nnU-Net  & MSD Spleen & 2& [1.60, 0.79, 0.79] & [64, 192, 160] & 0.2 & 1200  & 10 & 1\\ 
nnU-Net  & MSD Colon & 2& [3.0, 0.78, 0.78 ] & [56, 192, 192] & 0.2 & 1200  & 10 & 1\\ 

\bottomrule
\end{tabular}
\end{adjustbox}

\centering

\caption{The analysis of all tumor predictions larger than approximately $1cm^3$ on the test set shows the reduction of False Positive (FP) predictions using the proposed Region Loss. A clinician classified all tumor predictions to be True Positive (TP) if the prediction certainly indicates a tumor in the target region (e.g. a tumor within the target region liver).
Predictions within the corresponding target region, where a positive classification is not unambiguous, for instance, due to fluctuating image quality, are marked as True Region (TR). Predictions beyond the target region are always classified as FP.\label{tab_datasets_res}}
\begin{adjustbox}{width=0.8\textwidth}
\begin{tabular}{lcccc}
\toprule
Method & Class &  TR & TP & FP \\
\midrule
 \multirow{6}{*}{MultiTalent} & Liver Tumor &18 & 12& 23 \\
    &Lung Tumor & 6& 13& 25 \\
    &Pancreas Tumor & 13& 0& 19 \\
    &Hepatic Tumor & 14& 12& 24 \\
    \cmidrule(lr){2-5}
    &Sum & 51& 37& 91\\
    \midrule
    \midrule
\multirow{6}{*}{nnU-Net} & Liver Tumor&17 & 10& 10\\
    &Lung Tumor & 3&7 &8  \\
    &Pancreas Tumor & 9& 0& 0 \\
    &Hepatic Tumor & 7& 6& 19\\
    \cmidrule(lr){2-5}
    &Sum & 36& 23& 37\\

    \bottomrule
\end{tabular}%
  \quad%
\begin{tabular}{lcccc}
\toprule
Method & Class &  TR & TP & FP \\
\midrule
 \multirow{6}{*}{\makecell{MultiTalent \\ + \\ Region Loss \\ (proposed)}} & Liver Tumor &15 & 12& 1 \\
    &Lung Tumor & 3& 17& 7 \\
    &Pancreas Tumor & 11& 0& 2 \\
    &Hepatic Tumor & 15& 13& 3 \\
    \cmidrule(lr){2-5}
    &Sum & 44& 42& 13\\
    \midrule
    \midrule
\multirow{6}{*}{\makecell{nnU-Net \\ + \\ Region Loss \\ (proposed)}} & Liver Tumor&15& 12& 8\\
    &Lung Tumor & 3&10 &9  \\
    &Pancreas Tumor & 8& 0& 0 \\
    &Hepatic Tumor & 10& 5& 19\\
        \cmidrule(lr){2-5}
    &Sum & 36& 27& 36\\
    \bottomrule
\end{tabular}
\end{adjustbox}
\end{table}

\end{document}